\documentclass[preprint,authoryear,12pt]{elsarticle}

\usepackage{lineno,hyperref}
\usepackage{multicol}
\usepackage{graphicx}
\modulolinenumbers[5]

\begin{document}
\begin{frontmatter}

\title{Dwarf irregular galaxies with extended HI gas disks: Suppression of small-scale spiral structure by dark matter halo}
\author[mymainaddress]{Soumavo Ghosh\corref{mycorrespondingauthor}}
\cortext[mycorrespondingauthor]{Corresponding author}
\ead{soumavo@iisc.ac.in}

\author[mymainaddress,mysecondaryaddress]{Chanda J Jog}
\ead{cjjog@iisc.ac.in}

\address[mymainaddress]{Department of Physics, Indian Institute of Science,Bangalore 560012, India}

 \begin{abstract}
 Dwarf irregular galaxies with extended $HI$ disk distributions, such as DDO 154, allow measurement of rotation curves, hence
deduction of dark matter halo properties
to large radial distances, up to several times the optical radius.
 These galaxies contain a huge reservoir of dark matter halo, 
 which dominates over most of disk.
We study the effect of 
the dark matter halo on small-scale spiral features by
carrying out the local, non-axisymmetric perturbation analysis in the disks of five such late-type, gas-rich dwarf irregular galaxies, namely, DDO~154, NGC~3741, DDO~43, NGC~2366, and DDO~168 which host a dense and compact dark matter halo. 
We show that when the gas disk is treated alone, it allows a finite swing amplification; which would  result in 
small-scale spiral structure in the outer gas disk, but the addition of dark matter halo in the analysis results in a higher Toomre $Q$ parameter which prevents the amplification almost completely. This trend is also seen to be true in regions inside the optical radius. This implies absence of strong small-scale spiral arms in  these galaxies, which is in agreement with observations. 
Hence despite being gas-rich, and in fact having gas as the main baryonic component, these galaxies cannot support small-scale spiral structure  which would otherwise have been expected in normal gas-rich galaxies.

\end{abstract}

\end{frontmatter}
%

\section{Introduction} 
Dwarf galaxies are  the most common type of galaxies in the Local group  \citep{Mat98,THT09}.
They exhibit a huge variety of properties, from gas-less dwarf elliptical galaxies to gas-rich irregular galaxies.
Dwarf irregular galaxies form a subset of dwarf galaxies, and can be said to lie at the tail end of the Hubble sequence. They are thus late-type,
gas-rich, bulge-less galaxies. While these galaxies are still rotationally dominated, the maximum rotation velocity is small ($< 60$ km s$^{-1}$) \citep{Io16} as compared to values for normal high surface brightness (hereafter HSB) galaxies, like our Milky Way.

In this paper, we focus on late-type, gas-rich, rotationally supported dwarf galaxies, 
which host an extended $HI$ gas disk. In extreme cases, it can extend up to several times the optical radius. 
For example, the $HI$ disk is extended up to 4 to 8 times the Holmberg radius  in case of DDO 154 \citep{CF88} or NGC 3741 \citep{BCK05,Gen07}  respectively; thus allowing one to model the dark matter halo parameters up to large radii in these galaxies.
{\emph{ Over most of the disk, the $HI$ gas is the main constituent of the baryonic mass}.}
There is very little or no signature of molecular hydrogen (H$_2$) as traced by the $CO$ observations \citep{TaYo87,TKS87}, although it is also possible that the $CO$-to-H$_2$ conversion factor in such low metalicities is small, thus making $CO$ as a poor tracer for H$_2$ \citep{Ohta93,Buy06}.

Various surveys on galaxy morphology have shown that the spiral structure in disk galaxies can be divided into two types: small-scale, patchy or flocculent structure; or the more regular, grand-design spiral structure \citep[e.g. see][]{Elm11}.

Surprisingly having gas as the dominant baryonic component,
the late-type dwarf irregular galaxies with extended $HI$ disk lack strong spiral structure \citep{Ash92,Tol06}. This is counter-intuitive in the sense that any gas-rich, normal HSB galaxy would generally be expected to produce more flocculent spiral structure \citep[e.g. see][]{Jog92}.
It is well-known that the spiral arms take part in the secular evolution of disk galaxies by scattering the stars off the galactic plane \citep{BW67} and via angular momentum transport \citep{LYKA72,SJ14}. Thus a lack of spiral structure would imply that the secular evolution of the galactic disk is hampered.

\citet{KoFre16} found a systematic trend that the dark matter halos in dwarf galaxies have smaller core radii and denser core density as the galaxies become progressively less luminous. Following this trend, we expect that the dark matter halo would be more important from smaller radii for low luminosity dwarf galaxies that we study here. 
 Thus these galaxies could be good candidates for probing the effect of dark matter halo on the dynamics and evolution of the baryonic disks at the far end of the Hubble sequence.

Several past studies have investigated the dynamical role of a dominant dark matter halo in disk galaxies. \citet{Mih97} showed that the dominant dark matter halo prevents the global non-axisymmetric bar mode in the low surface brightness (hereafter LSB) galaxies. \citet{DeSe98} showed that in barred galaxies the dynamical friction from a dense dark matter halo slows down drastically the rotation speed of the bar. A dense, compact halo that dominates from the innermost regions as in LSB galaxies is shown to result in the superthin nature of some of these galaxies \citep{BJ13}, and also it suppresses swing amplification in these galaxies \citep{GJ14}. Even in a typical HSB galaxy like the Milky Way, the disk by itself is close to being unstable to local axisymmetric features, thus the dark matter halo is crucial in stabilizing such disks \citep{Jog14}.

Here, we carry out the dynamical study of local non-axisymmetric perturbations in the $HI$-rich dwarf irregular galaxies for which the required quantities such as $HI$ surface density, rotation curve have been measured observationally and are available in the literature. We choose five galaxies, namely, DDO~154, NGC~3741, DDO~43, NGC~2366 and DDO~168, and it turns out that each of these has a dense and compact dark matter halo. 
We then investigate the influence of dark matter halo on small-scale spiral structure as generated by swing amplification.

 We show that for all the five galaxies considered here, the dominant dark matter halo makes the gas disk rotationally more stable (as reflected from a high Toomre $Q$ value) which in turn prevents the finite growth of the non-axisymmetric perturbations in the disk almost completely.

\S~2 gives the details of sample of dwarf galaxies selected and the formulation of the problem; \S~3 describes the input parameters for different galaxies and the results while \S~4 and \S~5 contain the discussion and conclusions, respectively.

\section{Formulation of the problem}

\subsection{Sample of dwarf galaxies}

We select five late-type dwarf irregular galaxies, namely, DDO~154, NGC~3741, DDO~43, NGC~2366 and DDO~168 in which we study the constraining influence of dominant dark halo on the formation of small-scale spiral arms. Two of these (DDO~154, NGC~2366) are a part of the THINGS (The HI Nearby Galaxy Survey) survey by \citet{Wal08}, whereas DDO~43 and DDO~168 are a part of the LITTLE (Local Irregulars That Trace Luminosity Extremes) THINGS (The HI Nearby Galaxy Survey) survey by \citet{Hun12}. Our sample galaxies have the following characteristics:\\
 These have a large $HI$ disk (extending to several optical radii), and in the inner regions as well as in the outer regions the larger part of the baryonic contribution comes from $HI$ gas, and not stars. The galaxies are rotationally supported. We find that for these low luminosity galaxies, the corresponding dark matter halo is
dense and compact, i.e., the central density ($\rho_0$) is few times $10^{-2}$ M$_{\odot}$ pc$^{-2}$ and the core radius (R$_{\rm c}$) is less than two-three times the exponential disk scalelength (R$_{\rm d}$). Note that, the term `compact' does not imply anything about its extent.
 Thus for these dwarf galaxies, the dark matter halo dominates over most of the disk. In contrast, for HSB galaxies like the Milky Way and M~31, the core radius ($R_{\rm c}$) is of the order of 3--4 times the disk scalelength $R_{\rm d}$ \citep{BJ13}.

 Also as a counter-example, we consider IC~2574 (part of THINGS survey, \citep{Wal08}), a dwarf galaxy which has an extended $HI$ disk but the dark matter halo has low central density and is not compact (R$_{\rm c}>$ 3R$_{\rm d}$); a comparison with the above sample helps to bring out the  role of  a dominant dark matter halo on small-scale spiral features in $HI$-rich dwarf irregular galaxies.

\subsection{Non-axisymmetric perturbation : Swing amplification}

 The formulation of local, non-axisymmetric linear perturbation analysis of a galactic disk is followed from \citet{GLB65}. For the sake of completeness here we only mention the relevant assumptions and equations, for details see \citet{GLB65}.

The galactic disk is taken to be infinitesimally thin which is embedded in a dark matter halo, concentric to the galactic disk. For the sake of simplicity, here we treat the dark matter halo as rigid and non-responsive, i.e., it remains inert to the perturbations by the galactic disk.

The perturbations are taken to be planar, and the gas disk is taken to be isothermal, characterized by the surface density $\Sigma$ and the one-dimensional velocity dispersion or the sound speed $c$.

First we introduce sheared coordinates (to take account of the shear introduced by the differential rotation of the disk) given by:\\
\begin{eqnarray}
x'=x,\: y'=y-2Axt,\: z'=z, \: t'=t
\end{eqnarray}

Next we perform the linear perturbation analysis on the Euler's equations of motion, the continuity equation, and the Poisson equation, and a trial solution of the form exp[$i(k_x x' + k_yy')$] is introduced for the independent perturbed quantities such as the perturbed surface density $\delta \Sigma$.
We define $\tau$ as:\\
\begin{equation}
\tau \equiv 2At'-k_x/k_y \hspace{0.3 cm} \mbox{, for a wavenumber} \hspace{0.1 cm} k_y \ne 0
\end{equation}
In the sheared coordinates, $\tau$ is a measure of time, and it becomes zero when the modes becomes radial, i.e., where $x$ is along the initial radial direction.

We define $\theta$, the dimensionless measure of the density perturbation as:\\
\begin{equation}
\theta = \delta \Sigma/ \Sigma
\end{equation}
\noindent where $\Sigma$ denotes the unperturbed surface density and  $\delta \Sigma$ denotes the variation in surface density in the sheared frame whereas in the non-sheared galactocentric frame of reference, it denotes the density for a mode of wavenumber $k_y(1+\tau^2)^{1/2}$ that is sheared by an angle $\tan^{-1}\tau$ with respect to the radial position. It implies that higher the value of $\tau$, the more sheared will be the mode.

After some algebraic simplifications along with the usage of equations (2) and (3), the perturbed equations of motion, the continuity equation, and the Poisson equation reduce to \citep[for details see][]{GLB65}
\begin{equation}
\left(\frac{d^2\theta}{d\tau^2}\right)-\left(\frac{d\theta}{d\tau}\right)\left(\frac{2\tau}{1+\tau^2}\right)+\theta\Bigg[\frac{\kappa^2}{4A^2}+\frac{2B/A}{1+\tau^2}\\
+\frac{k_y^2c^2}{4A^2}(1+\tau^2)-\Sigma\left(\frac{\pi G k_y}{2A^2}\right)(1+\tau^2)^{1/2}\Bigg]=0\,.
\label{swing-2comp}
\end{equation}
\noindent where $\kappa$ is the local epicyclic frequency and $A$, $B$ are the Oort constants.

The four terms within the square brackets of equation (\ref{swing-2comp}) are due to the epicyclic motion, the unperturbed shear flow, the gas pressure, and the self-gravity, respectively.

 The systematic behavior of $\theta$ is as follows:

 When $\tau$ is large, the pressure term dominates over other terms, and hence the solution will be oscillatory in nature, but when $\tau$ is small the epicyclic motion term and the unperturbed shear flow dominate over the pressure term and they cancel each other completely for a flat rotation curve, thus resulting in setting up a kinematic resonance. In addition, if the self-gravity term dominates over the pressure term then the duration of kinematic resonance increases and the mode undergoes a swing amplification while evolving from radial position ($\tau =0$) to trailing position ($\tau > 0$) \citep[for details see][]{GLB65,Too81}.

Now we introduce three dimensionless parameters, namely, Toomre $Q$ parameter \citep{Too64} = $\kappa c / \pi G \Sigma$ where  $c$ is the velocity dispersion, $\eta$ (= $2A/\Omega$) which denotes the logarithmic shearing rate and $X$ = ($\lambda_y/\lambda_{\rm crit}$), where $\lambda_{\rm crit} (= 4\pi^2 G \Sigma/\kappa^2)$ is the critical wavelength for growth of instabilities in a one-fluid disk supported purely by rotation.

Putting these quantities in equation~(\ref{swing-2comp}), we finally get the evolution of $\theta$ with $\tau$ as
\begin{equation}
\left(\frac{d^2\theta}{d \tau^2}
\right)-\left(\frac{d\theta}{d\tau}\right)\left(\frac{2\tau}{1+\tau^2}
\right)+\theta\Big[{\xi^2}+\frac{2(\eta-2)}{\eta(1+\tau^2)}\\
+\frac{(1+\tau^2)Q^2\xi^2}{4X^2}-\frac{\xi^2}{X}(1+\tau^2)^{1/2}\Big]=0\,,
\label{swing-final}
\end{equation}
\noindent where ${\xi^2}={\kappa^2/4A^2}= 2(2-\eta)/\eta^2$.

For a given set of parameter values we solve equation~(\ref{swing-final}) numerically by fourth-order Runge-Kutta method while treating equation~(\ref{swing-final}) as two coupled, first--order linear differential equations in $\theta$ and $d\theta/d\tau$ \citep[for details see][]{Jog92}. We point out that for this linear analysis, the ratio of perturbed to unperturbed gas surface density ($\theta_{\rm g}$) may be multiplied by a arbitrary scale factor (say $\alpha$) such that the net fractional amplitude $\alpha \theta_{\rm g}$ remains $\ll 1$, for all $\tau$ values considered \citep[for details see][]{Jog92}.

\section{Results}
In this section we present results of the swing amplification at different radii for the five galaxies considered here. First we give the sources from which we obtained the observed values of the required parameters such as the rotation curve, surface density etc.; as well as the halo model parameters. These values  are listed in Table~1 (for detailed description see text in \S~3.1). For the sake of consistency, we choose the halo parameters which are obtained by fitting a pseudo-isothermal density profile as given in the cited references. 

 We clarify that for any individual galaxy, although various parameters for stellar, gas and dark matter halo are taken from different papers (as given below), the observational data-set used in those papers are the same. Therefore, these parameters are internally consistent for each individual galaxy. For example, \citet{deB08} gives the values of dark matter halo parameters and \citet{Ler08} gives the values of stellar parameters for DDO~154, although both of them use the same THINGS data.

\subsection{Input parameters}
{\emph{DDO~154}}: The rotation curve 
is taken from \citet{deB08}. The stellar parameters $\Sigma_0$ (central surface density) and exponential scalelength (R$_{\rm d}$) are taken from \citet{Ler08} whereas the dark matter halo parameters are taken from \citet{deB08}. We use the $HI$ surface density distribution and a constant value for the gas velocity dispersion (8 km s$^{-1}$) from the maps given in \citet{Wal08}. The extent of $HI$ disk is taken from \citet{Bage11}.

{\emph{NGC~3741}}: The rotation curve, stellar parameters and the dark matter halo parameters are taken from \citet{BCK05} whereas the gas mass and the gas velocity dispersion value are taken from \citet{Beg13}. The extent of $HI$ disk is taken from \citet{BCK05}.

{\emph{DDO~43}}: The rotation curve is taken from \citet{Oh15}. The disk scalelength is from \citet{HE04}. The $HI$ surface density, surface density and the total amount of $HI$ gas and the dark matter halo parameters are given by \citet{Oh15}. Note that, since \citet{Oh15} did not consider stellar component (as it is very small) and carried out the mass modeling from the rotation curve, therefore, for internal consistency, we do not consider the stellar component. We assumed a constant $HI$ velocity dispersion of 8 km s$^{-1}$, typical of what we found for other dwarf galaxies in this sample. The extent of $HI$ disk is taken from \citet{SHN05}.

{\emph{NGC~2366}}: The rotation curve and the stellar parameters ($\Sigma_0$, R$_{\rm d}$) are taken from \citet{Oh08} whereas the dark matter halo parameters are from \citet{deB08}. The $HI$ surface density distribution and a constant value for the gas velocity dispersion (9 km s$^{-1}$) are taken from the maps given in \citet{Wal08}. The extent of $HI$ disk is taken from \citet{Bage11}.

{\emph{DDO~168}}: The rotation curve, the component-wise decomposition, stellar mass and the average gas velocity dispersion are taken from \citet{John15} whereas the disk scalelength is taken from \citet{HE04}. The dark matter halo parameters are taken from \citet{Oh15}. The $HI$ surface density is taken from LITTLE THINGS by \citet{Hun12}. The extent of $HI$ disk is taken from \citet{Bro92}.

{\emph{IC~2574}}: The rotation curve is taken from \citet{Oh08}. We take the stellar parameters from \citet{Ler08} and the dark matter halo parameters from \citet{deB08}. The $HI$ surface density distribution and a constant value for the gas velocity dispersion (7 km s$^{-1}$) are taken from the maps given in \citet{Wal08}. The extent of $HI$ disk is taken from \citet{Bage11}.

There are other isolated, dwarf galaxies, e.g. DDO~170 \citep{LSV90}, DDO~46 \citep{Oh15} which could be good candidates for our study since these have dense and compact halo, however all the parameters are not known observationally; hence
this restricts us to consider the above sample of five galaxies.

\begin{table*}
\begin{minipage}{\textwidth}
   \caption{Input parameters for the sample galaxies}
\begin{tabular}{cccccccccc}
\hline
galaxy & RC ref.& $\Sigma_{\rm star}$ & $M_{\rm gas}$ & R$_{\rm d}$ & $R_{\rm HI}/ R_{\rm d}$ & $\rho_0$ & R$_{\rm c}$\\
& & (M$_{\odot}$ pc$^{-2}$) & ($\times$10$^8$ M$_{\odot}$)& (kpc) && (M$_{\odot}$ pc $^{-3}$) & (kpc)\\
\hline
DDO~154 &(1) & 5.7$^{a}$ & 3.6$^{e}$& 0.8$^{a}$ & 8.5$^{j}$ & 0.028$^{b}$ & 1.34$^{b}$\\
NGC~3741 &(2) & 2.7$^{d}$ & 1.7$^{c}$& 0.9$^{d}$ & 13.1$^{d}$ & 0.078$^{d}$ & 0.7$^{d}$ \\
DDO~43 & (4) & - & 2.3$^{h}$ & 0.43$^{f}$ & 9.1$^{m}$ & 0.033$^{h}$ & 0.94$^{h}$\\
NGC~2366 & (3) & 10.5$^{b}$ & 6.5$^{e}$& 0.5$^{b}$ & 13.6$^{j}$ & 0.035$^{b}$ & 1.36$^{b}$ \\
DDO~168 & (5) & 21.2$^{i}$ & 2.6$^{g}$& 0.66$^{h}$ & 11.7$^{k}$ & 0.039$^{h}$ & 2.8$^{h}$ & \\
$*$IC~2574 & (3) & 24.6$^{a}$ & 14.8$^{e}$& 2.1$^{a}$ & 4.4$^{j}$ & 0.004$^{b}$ & 7.23$^{b}$ \\
\hline
\end{tabular}
{ $\Sigma_{\rm star}$ denotes the central stellar surface density and $M_{\rm gas}$ denotes the total gas mass of the galaxy. $R_{\rm HI}$ and $R_{\rm d}$ denote the extent of the $HI$ disk and the disk scalelength, respectively. $\rho_0$ is the core density and $R_{\rm c}$ is the core radius of the dark matter halo.\\
$*$IC~2574 is used as an counter-example, as it has a non-dense and non-compact dark matter halo.\\
Rotation curve (RC) refs: (1)-\citet{deB08};(2)-\citet{BCK05}; (3)-\citet{Oh08}; (4)-\citet{Oh15}; (5)-\citet{John15}.\\
Other parameter refs.: (a)-\citet{Ler08}; (b)-\citet{deB08}; (c)-\citet{Beg13}; (d)-\citet{BCK05}; (e)-\citet{Wal08}; (f)-\citet{HE04};
 (g)-\citet{Hun12};(h)-\citet{Oh15};(i)-\citet{John15}; (j)-\citet{Bage11}; (k)-\citet{Bro92}; (l)-\citet{Oh08}; (m)-\citet{SHN05}.}
\end{minipage}
\end{table*}

\subsection{Effect of dark matter halo : suppression of swing amplification}

 Before going into the details of the result we point out that for the radii considered here, the rotation curve is assumed to be flat and thus $\eta$ = 1 is used for all cases. The dependence of our finding with different values of $\eta$ is studied later in this paper (see \S~3.3 ). We use $X$ = 2 for all cases as in the past literature, it was shown that the swing amplification reaches its peak at around $X=2$ \citep[e.g. see][]{Too81,BT87}. 

 Since the disk scalelength is different for these galaxies, and the observational data-points for the rotation curve and surface densities of stellar and gas are available for different ranges of radii, therefore maintaining exact uniformity in selecting radii is 
not possible. 
We point out that, at the radii chosen, the larger contribution to the baryonic mass is due to $HI$ and not stars. Hence, the one-component treatment outlined in \S~2.2 is valid.

Also note that we have used the stellar disk scalelength, R$_{\rm d}$, as a measure of spatial size as is the usual practice in dynamical studies of galaxies, even though the main baryonic component of the disk is $HI$ gas. Similarly, following the
usual practice, we have given $HI$ extent in terms of stellar disk scalelength, R$_{\rm d}$ (Table~1).

\subsubsection{DDO~154}
Keeping  the criteria as given in \S~3.2, we choose three radii, namely, 4 R$_{\rm d}$, 5 R$_{\rm d}$, and 6 R$_{\rm d}$. At these radii, the surface density values for $HI$ ($\Sigma_{HI}$) are 3.9, 3, and 2.1 M$_{\odot}$ pc$^{-2}$, respectively \citep{Wal08}. The Holmberg radius ($R_{\rm Ho}$) for this galaxy is 2.1 $R_{\rm d}$.
 
\begin{figure}
    \centering
    \begin{minipage}{.5\textwidth}
        \centering
        \includegraphics[height=2.5in,width=3.5in]{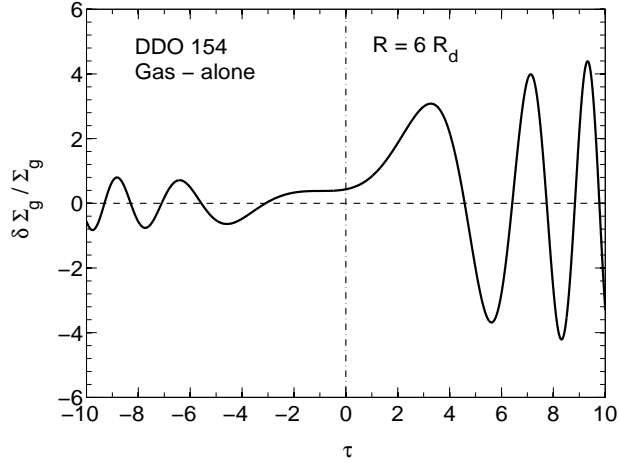}
       \vspace{0.2 cm}
	{\bf{(a)}}\\
    \end{minipage}
\begin{minipage}{.5\textwidth}
        \centering
        \includegraphics[height=2.5in,width=3.5in]{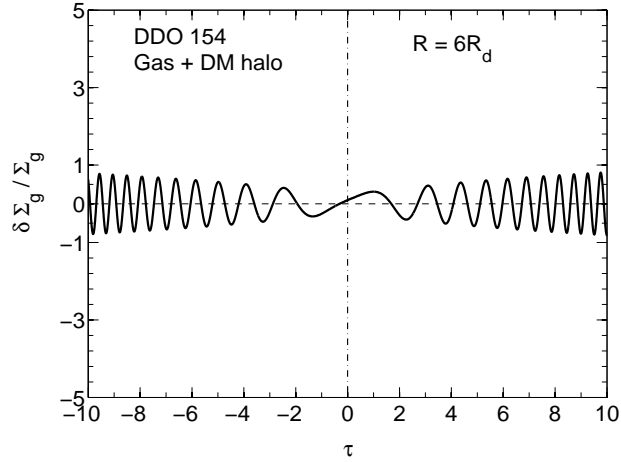}
        \vspace{0.2 cm}
	{\bf{(b)}}\\
    \end{minipage}
    \caption{DDO~154: Variation in $\delta \Sigma_{\rm g}/\Sigma_{\rm g}$, the ratio of the perturbed gas surface density to the unperturbed gas surface density plotted as a function of $\tau$, dimensionless measure of time in the sheared frame, at $R$ = 6 R$_{\rm d}$. (a)  For the gas-alone case ($Q = 1.3$) while (b) is for the gas plus dark matter halo case ($Q = 3.8$). While the gas-alone case shows finite amplification, the inclusion of the dark matter prevents the amplification, which would thus prevent the occurrence of strong small-scale spiral features. The net fractional amplitude $\alpha \theta_{\rm g}$ is $\ll 1$ at all $\tau$, where $\alpha$ is a scale factor.}
    \label{DDO154_first}
\end{figure}

Now we investigate the role of dominant dark matter halo using the following prescription:\\
 Using the rotation curve calculated from observed gas surface density values (as done in e.g. \citet{CF88}) and assuming a flat rotation curve (as mentioned in \S~3.2) we calculate $\kappa$ for gas-alone case and then use it to calculate the Toomre $Q$ value.
 Then we take the observed net rotation curve (also being flat, but with a higher value for the rotation velocity than for gas-alone case) which also includes the contribution of the dark matter halo \citep[see][]{deB08} to calculate $\kappa$ and the corresponding Toomre $Q$ value. Then for these two Toomre $Q$ values, we separately solve equation~(\ref{swing-final}) for these two cases. 
This procedure will tell us whether dark matter halo plays a significant role in suppressing  small-scale spiral features generated through the swing amplification mechanism. We choose 6 R$_{\rm d}$ for this detailed analysis and we find that $Q$ = 1.3 when the gas-alone contribution is taken and $Q$ = 3.8 when the net rotation curve is used for calculating $\kappa$. The resulting swing amplification is shown in Fig.~\ref{DDO154_first}.

From Fig.~\ref{DDO154_first} it is clear that when only  gas in the disk is considered, the system shows a finite swing amplification; which indicates that small-scale spiral structure is possible in the gas-alone system, but when we also include the dark matter halo in the system, the net $Q$ value is higher, and this almost completely suppresses the swing amplification, thus preventing the system from displaying small-scale spiral features.
For the other radii, namely, $R= 4R_{\rm d}$ and $R=5R_{\rm d}$, we got a similar behavior, i.e., the gas-alone system allows a finite growth of the non-axisymmetric perturbations, but the addition of dark matter halo prevents the growth of the perturbations. Therefore, for these two radii, we do not show any figure.

Thus the suppression of swing amplification and hence strong small-scale spiral arms by the dominant dark matter halo turns out to be a general result for the radii we considered for DDO~154. 
 This trend also continues well inside the optical radius (see \S~3.4).

\subsubsection{NGC~3741}

We choose four radii, namely, 2.5 R$_{\rm d}$, 3 R$_{\rm d}$, 3.5 R$_{\rm d}$, and 4 R$_{\rm d}$. At these radii, the surface density values for $HI$ ($\Sigma_{HI}$) are 2.5, 1.5, 1, 0.5 M$_{\odot}$ pc$^{-2}$, respectively \citep{Beg13}. Note that the rotation curve of $HI$ disk is measured up to 8.3 R$_{\rm d}$ \citep[see fig. 1 in][]{BCK05}, but the $HI$ surface density data is limited to 4.5 $R_{\rm d}$ \citep[see fig. 5 in][]{Beg13}, thus restricting us to the above range chosen. The Holmberg radius ($R_{\rm Ho}$) for this galaxy is 1.6 $R_{\rm d}$.

\begin{figure}
    \centering
    \begin{minipage}{.5\textwidth}
        \centering
        \includegraphics[height=2.5in,width=3.5in]{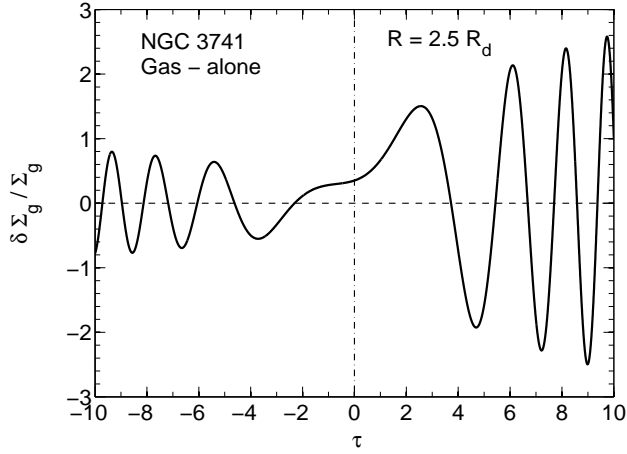}
       \vspace{0.2 cm}
	{\bf{(a)}}\\
    \end{minipage}
\begin{minipage}{.5\textwidth}
        \centering
        \includegraphics[height=2.5in,width=3.5in]{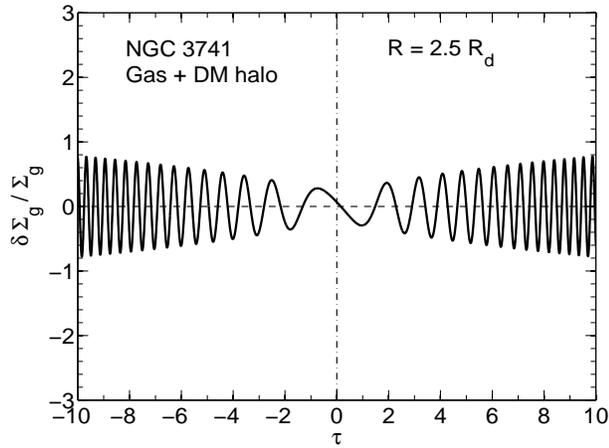}
        \vspace{0.2 cm}
	{\bf{(b)}}\\
    \end{minipage}
	\caption{NGC~3741: Variation in $\delta \Sigma_{\rm g}/\Sigma_{\rm g}$, the ratio of the perturbed gas surface density to the unperturbed gas surface density plotted as a function of $\tau$, dimensionless measure of time in the sheared frame, at $R$ = 2.5 R$_{\rm d}$. (a)  For the gas-alone case ($Q = 1.5$) while (b) is for the gas plus dark matter halo case ($Q = 5.3$). While the gas-alone case shows finite amplification, the inclusion of the dark matter halo prevents the amplification, which would thus suppress the occurrence of strong small-scale spiral features. The net fractional amplitude $\alpha \theta_{\rm g}$ is $\ll 1$ at all $\tau$, where $\alpha$ is a scale factor.}
    \label{fig-ngc3741}
\end{figure}

At $R=$ 2.5 R$_{\rm d}$, when the contribution of only gas to the rotation curve is taken into account, the Toomre $Q$ value is calculated to be 1.5, whereas by considering the net rotation curve we got the Toomre $Q$ value as 5.3. The resulting swing amplification is shown in Fig.~\ref{fig-ngc3741}.

From Fig.~\ref{fig-ngc3741} we see that gas-alone case allows a finite amplification in the perturbed gas surface density, but the addition of dark matter halo suppresses the amplification almost completely, thus indicating that small-scale spiral structure will be suppressed.  For other three radii, namely 3 R$_{\rm d}$, 3.5 R$_{\rm d}$, and 4 R$_{\rm d}$, when the net rotation curve (which represents the real galaxy) is used, the Toomre $Q$ values are found to be even higher, namely 8.3, 10.9 and 19.7, respectively. We checked that the values are too high to produce finite swing amplification. Therefore, dark matter halo is found to have a pivotal role in suppressing the small-scale spiral features at radii we considered for this galaxy. However, a weak spiral arm can be permitted in the disk (for details see discussion in \S~4.2). This trend also continues well inside the optical radius (see \S~3.4).

\subsubsection{DDO~43}

We choose three radii, namely, 6 R$_{\rm d}$, 7 R$_{\rm d}$, and 8 R$_{\rm d}$. At these radii, the surface density values for $HI$ ($\Sigma_{HI}$) are 5, 3, and 2.5 M$_{\odot}$ pc$^{-2}$, respectively \citep{Oh15}. The Holmberg radius ($R_{\rm Ho}$) for this galaxy is 4.1 $R_{\rm d}$.

\begin{figure}
    \centering
    \begin{minipage}{.5\textwidth}
        \centering
        \includegraphics[height=2.5in,width=3.5in]{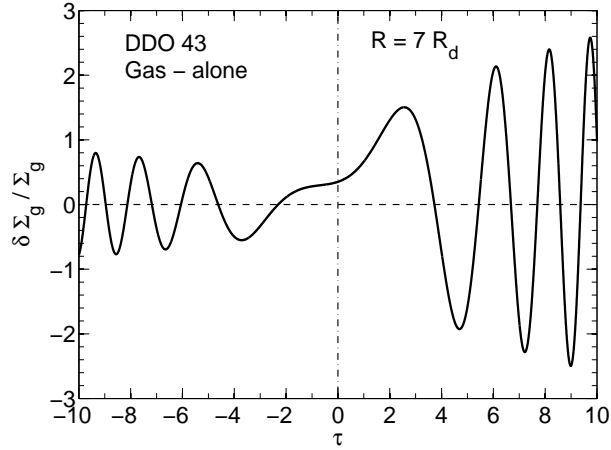}
       \vspace{0.2 cm}
	{\bf{(a)}}\\
    \end{minipage}
\begin{minipage}{.5\textwidth}
        \centering
        \includegraphics[height=2.5in,width=3.5in]{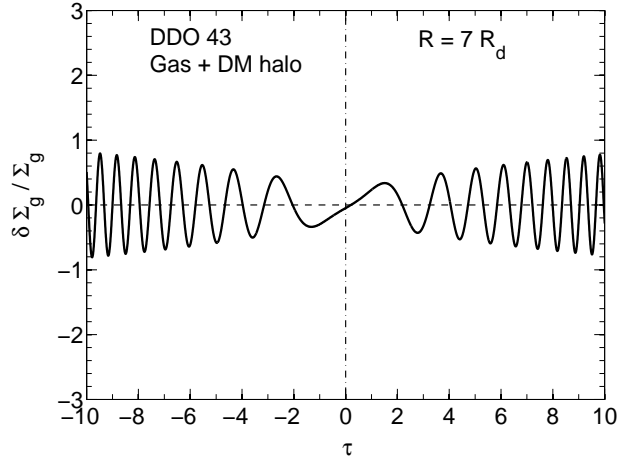}
        \vspace{0.2 cm}
	{\bf{(b)}}\\
    \end{minipage}
	\caption{DDO~43: Variation in $\delta \Sigma_{\rm g}/\Sigma_{\rm g}$, the ratio of the perturbed gas surface density to the unperturbed gas surface density plotted as a function of $\tau$, dimensionless measure of time in the sheared frame, at $R$ = 7 R$_{\rm d}$. (a)  For the gas-alone case ($Q = 1.5$) while (b) is for the gas plus dark matter halo case ($Q = 3.1$). While the gas-alone case shows finite amplification, the inclusion of dark matter halo prevents the amplification, which would thus suppress the occurrence of strong small-scale spiral features. The net fractional amplitude $\alpha \theta_{\rm g}$ is $\ll 1$ at all $\tau$, where $\alpha$ is a scale factor.}
    \label{fig-ddo43}
\end{figure}

 Following the same technique as mentioned in \S~3.2.1, we calculated the Toomre $Q$ parameter, first for the gas-alone case, and then for the gas plus dark matter halo case. For illustration purpose, we present the analysis for $R=7 R_{\rm d}$. At this radius, we solve equation~(\ref{swing-final}) by using the Toomre $Q$ parameter for both gas-alone and gas plus dark matter halo cases. This is shown in Fig.~\ref{fig-ddo43}. We find that, the gas-alone case allows finite swing amplification, but when the contribution of dark matter is taken into account, it completely damps the amplification that was earlier present in the gas disk.

For the other radii, namely, 6 R$_{\rm d}$ and 8 R$_{\rm d}$, we found the same trend as was seen for 7 R$_{\rm d}$, therefore, we do not present any figure for these cases. Hence, for this galaxy also, the dominant dark matter halo prevents the small-scale spiral structure at radii considered here. The role of dark matter on preventing the small-scale spiral arms in the regions well inside the optical disk is discussed in \S~3.4.

\subsubsection{NGC~2366}

We choose four radii, namely, 6 R$_{\rm d}$, 7 R$_{\rm d}$, 8 R$_{\rm d}$ and 10 R$_{\rm d}$. At these radii, the surface density values for $HI$ ($\Sigma_{HI}$) are 7.6, 6.7, 4.9, and 2.8 M$_{\odot}$ pc$^{-2}$, respectively \citep{Wal08}. For all these radii we obtained the values of Toomre $Q$ parameter, and they are 2, 1.9, 2.4, and 3.4, respectively, by using the net observed rotation curve which includes the contributions of gas and dark matter halo. We did not find finite swing amplification at these four radii considered here. Thus dark matter turns out to be the key factor in preventing the small-scale spiral structure.

\subsubsection{DDO~168}
We choose three radii, namely, 3.5 R$_{\rm d}$, 4 R$_{\rm d}$, and 5 R$_{\rm d}$. The $HI$ surface density ($\Sigma_{HI}$) at these radii are 2.6, 1.3, and 1 M$_{\odot}$ pc$^{-2}$, respectively \citep{Hun12}. The Holmberg radius ($R_{\rm Ho}$) for this galaxy is 4.2 $R_{\rm d}$.

The values of Toomre $Q$ parameter for gas-alone case are calculated to be 3.7, 6.5, and 6.6, respectively. Then we calculated the values of Toomre $Q$ parameter for the gas plus dark matter halo case, and the values are 11.4, 20.4, and 19.8, respectively.

  We see that the Toomre $Q$ values are very high in both the gas-alone and gas plus dark matter halo cases as compared to the other galaxies. Note that, for this galaxy, the value of $\kappa$ is higher than the other cases and the $HI$ velocity dispersion (11 km s$^{-1}$) is also a bit higher (~7-8 km s$^{-1}$ being typical values) than the other galaxies. Also, the $HI$ surface density is lower than the other galaxies considered. Thus all factors comprise to produce  very high values of Toomre $Q$ parameter. We did not find any amplification at all at these radii. While the gas-alone case does not allow any swing amplification, the inclusion of dark matter in the system completely rules out any possibility for the system to host small-scale spiral structure at radii that we considered here.  
The role of dark matter and the low disk surface density on preventing the small-scale spiral features in the regions well inside the optical disk is discussed in \S~3.4.

\subsection {Variation with $\eta$ -- departure from the assumption of flat rotation curve}

So far we have considered only flat rotation curve which corresponds to $\eta = 1$. We note that for the galaxies considered here, sometimes the rotation curves are not strictly flat in the regions where we have carried out the analysis of swing amplification, e.g., see the rotation curve of DDO~43 in \citet{Oh15} which is not strictly flat, but is slowly rising at radii which we have considered here. We calculated the actual value of $\eta$ from the observed rotation curve of DDO~43 at $R=7 R_{\rm d}$, and the value of $\eta$ turned out to be $\sim$ 0.6. This indicates a departure from the flat rotation curve assumption.

 Here in this section, we study the dependence of our finding on the variation of the parameter $\eta$.

The parameter $\eta$ as defined earlier in \S~2.2 can be expressed in detail as 
\begin{equation}
\eta = \frac{2A}{\Omega} = \frac{1}{\Omega}\left[\frac{v_{\rm c}}{R} - \frac{dv_{\rm c}}{dR}\right]
\end{equation}
\noindent where $v_{\rm c}$ denotes the circular velocity at radius $R$.

\begin{figure}
    \centering
    \begin{minipage}{.5\textwidth}
        \centering
        \includegraphics[height=2.5in,width=3.5in]{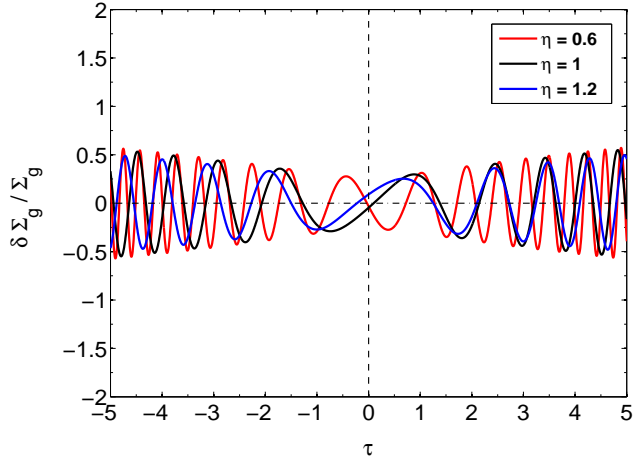}
    \end{minipage}
	\caption{DDO~43: The evolution of $\delta \Sigma_{\rm g}/\Sigma_{\rm g}$, the ratio of the perturbed gas surface density to the unperturbed gas surface density plotted as a function of $\tau$, dimensionless measure of time in the sheared frame at $R$ = 7 R$_{\rm d}$ ($Q$ = 3.1), with $\eta = 1.2$ and $0.6$, along with the standard case of $\eta = 1$. The suppression of small-scale spiral features by dominant dark matter halo remains unchanged with the variation of $\eta$ (for details see text).}
    \label{fig-variation_eta}
\end{figure}

Therefore, $\eta$ $> < 1$ would imply a falling rotation curve and a rising rotation curve, respectively. For illustration purpose we choose $R=7 R_{\rm d}$ of DDO~43 where the Toomre $Q$ value for gas plus dark matter halo system is found to be 3.1. Then we carry out the swing amplification analysis with $\eta =0.6,0.8, 1.2$. The results for swing amplifications for $\eta = 1.2$ and $\eta = 0.6$ , along with $\eta =1$ are shown in Fig~\ref{fig-variation_eta}.

From Fig.~\ref{fig-variation_eta} it is clear that for $\eta = 1.2$ and $\eta = 0.6$, the corresponding solutions obtained from equation (\ref{swing-final}) remains indistinguishable from that obtained for $\eta = 1$. Therefore, even if the rotation curve is either slightly falling or rising, the main finding of this paper remains unchanged. In other words, if for a radius $R$ and for $\eta = 1$, the dominant dark matter halo is found to suppress the small-scale spiral features, the same will also hold for $\eta > 1$ and $\eta < 1$.

The general trend of change in swing amplification for different $\eta$ values is as follows:\\
For $\eta$ $< 1$ (i.e. for rising rotation curve), the pressure term stops the growth of the perturbations at an early epoch and hence would result in more open spiral structure, and for $\eta >1$ (i.e. for falling rotation curve) the effect will be opposite to the case for $\eta <1$ \citep[e.g., see][]{GLB65,Jog92}.

Here, the Toomre $Q$ value is so high that the system will not be able to support any swing amplification, and hence the issue of relative openness of the spiral structure (as generated by swing amplification) will appear for this work.  Thus, variation of $\eta$ has little impact on the main finding of this paper. The high Toomre Q values effectively damp swing amplification for any of the $\eta$ values considered here.

\subsection { Suppression of spiral structure at all radii-- a generic trend ?}

 So far, we showed the effect of dark matter halo on suppression of the strong small-scale spiral features at some radii for each galaxy considered. 
The set of radii we chose for the calculation of swing amplification lie all outside the optical disk of the respective galaxies. 
Hence, the question remains whether this effect of dark matter halo on prevention of strong small-scale spiral structure is strictly valid only in the outer disks or this trend also holds true for regions well inside the optical disk? 

\begin{figure*}
\begin{multicols}{2}
    \includegraphics[width=\linewidth]{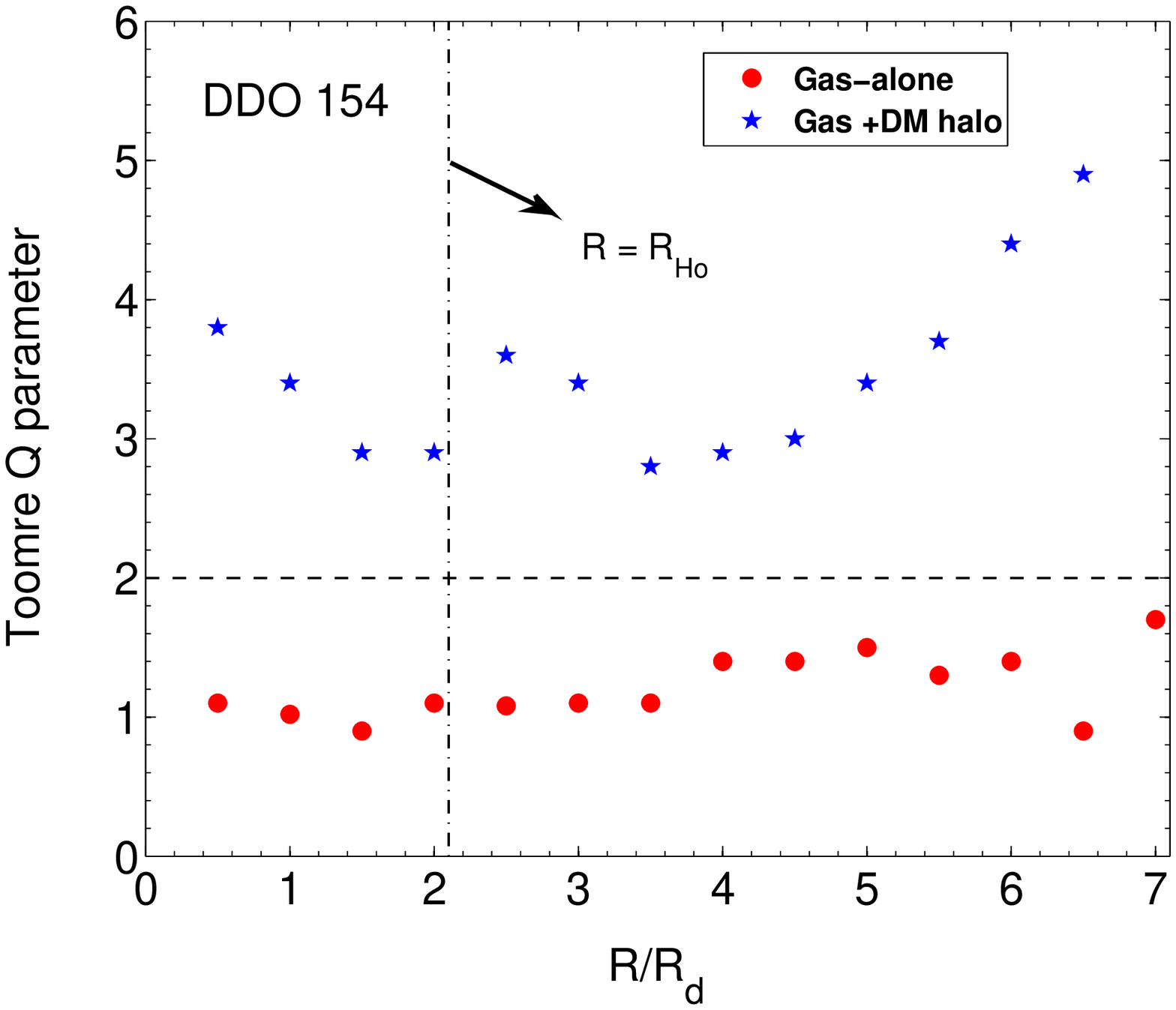}\par 
    \includegraphics[width=\linewidth]{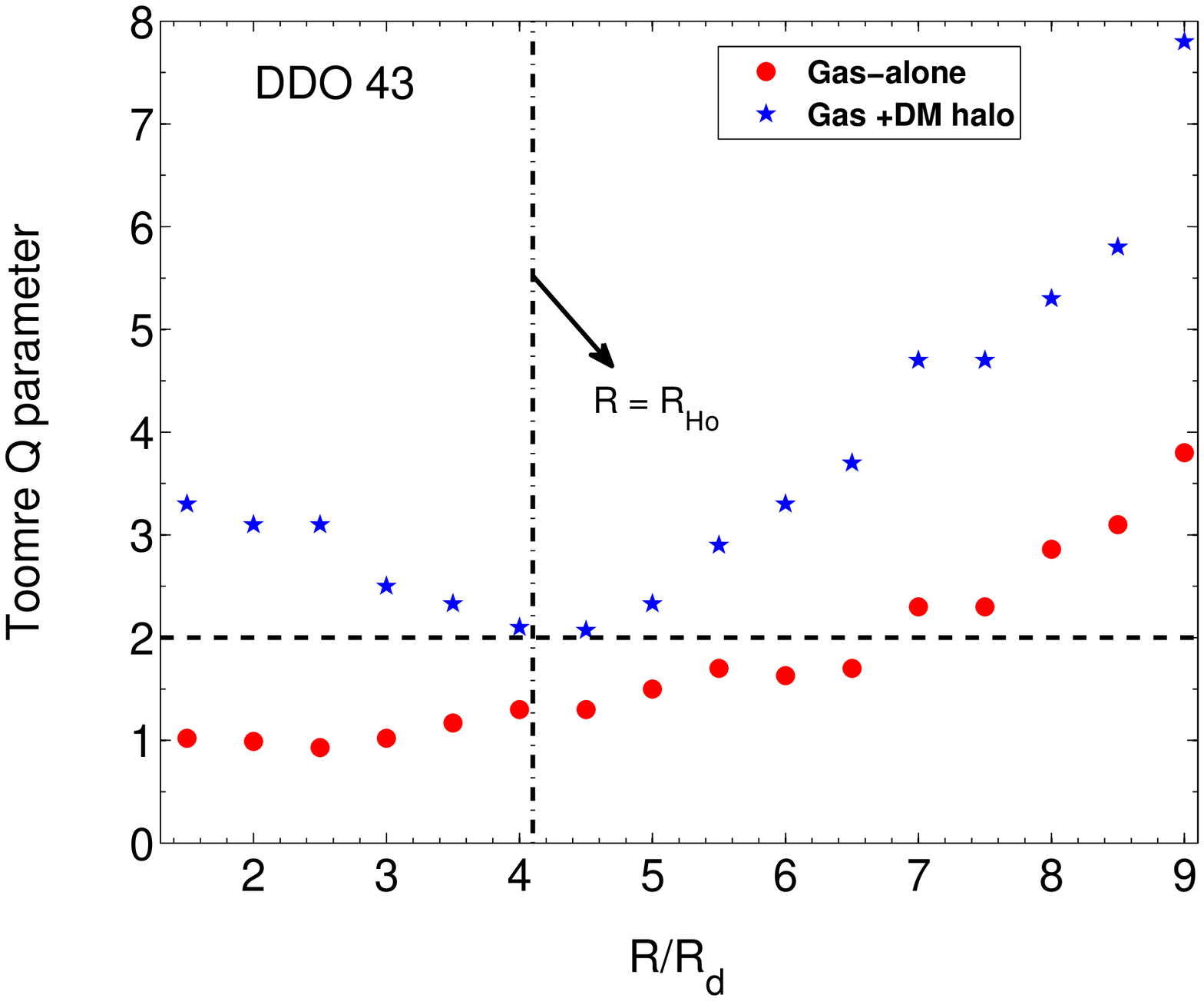}\par 
    \end{multicols}
\begin{multicols}{2}
    \includegraphics[width=\linewidth]{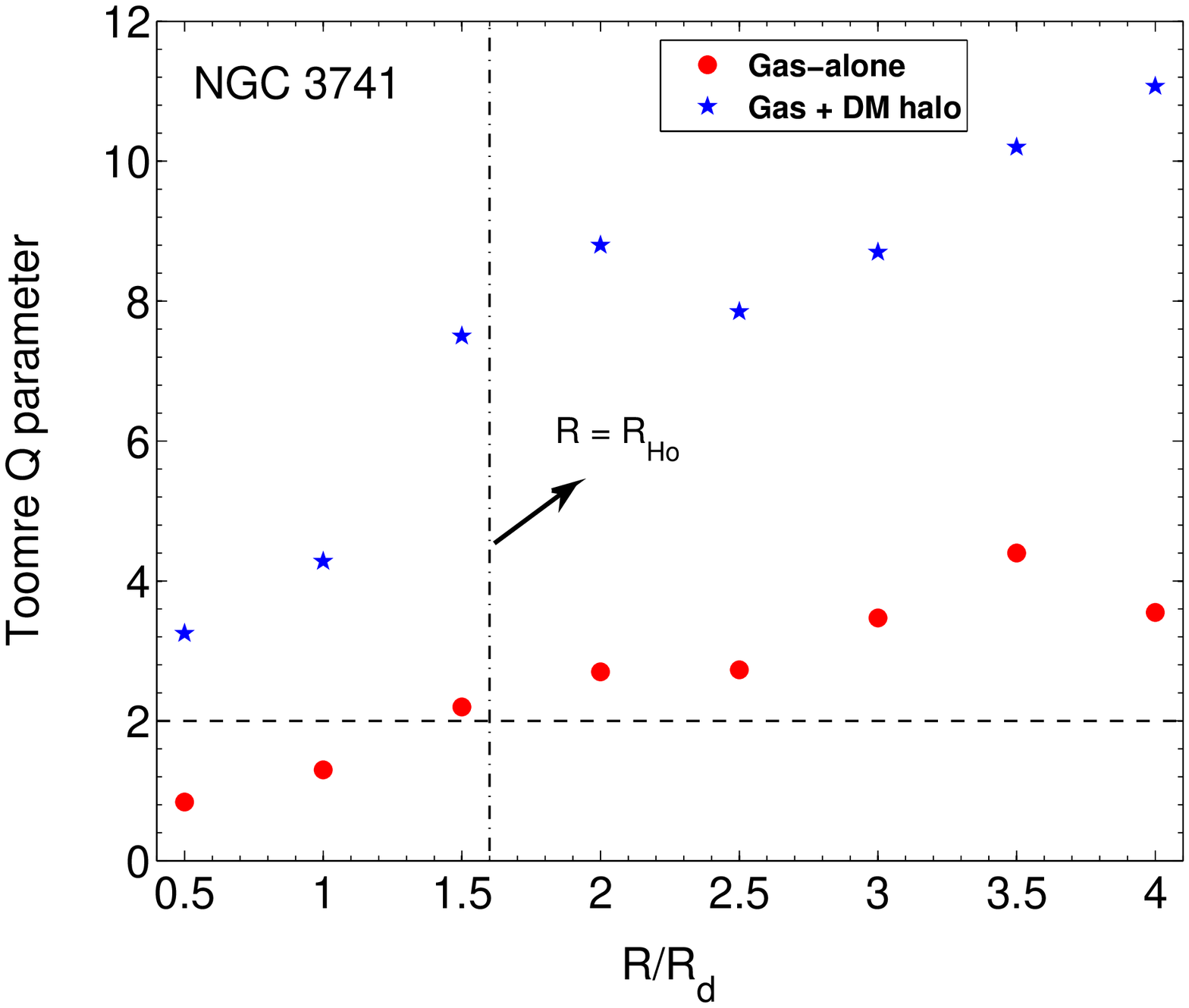}\par
    \includegraphics[width=\linewidth]{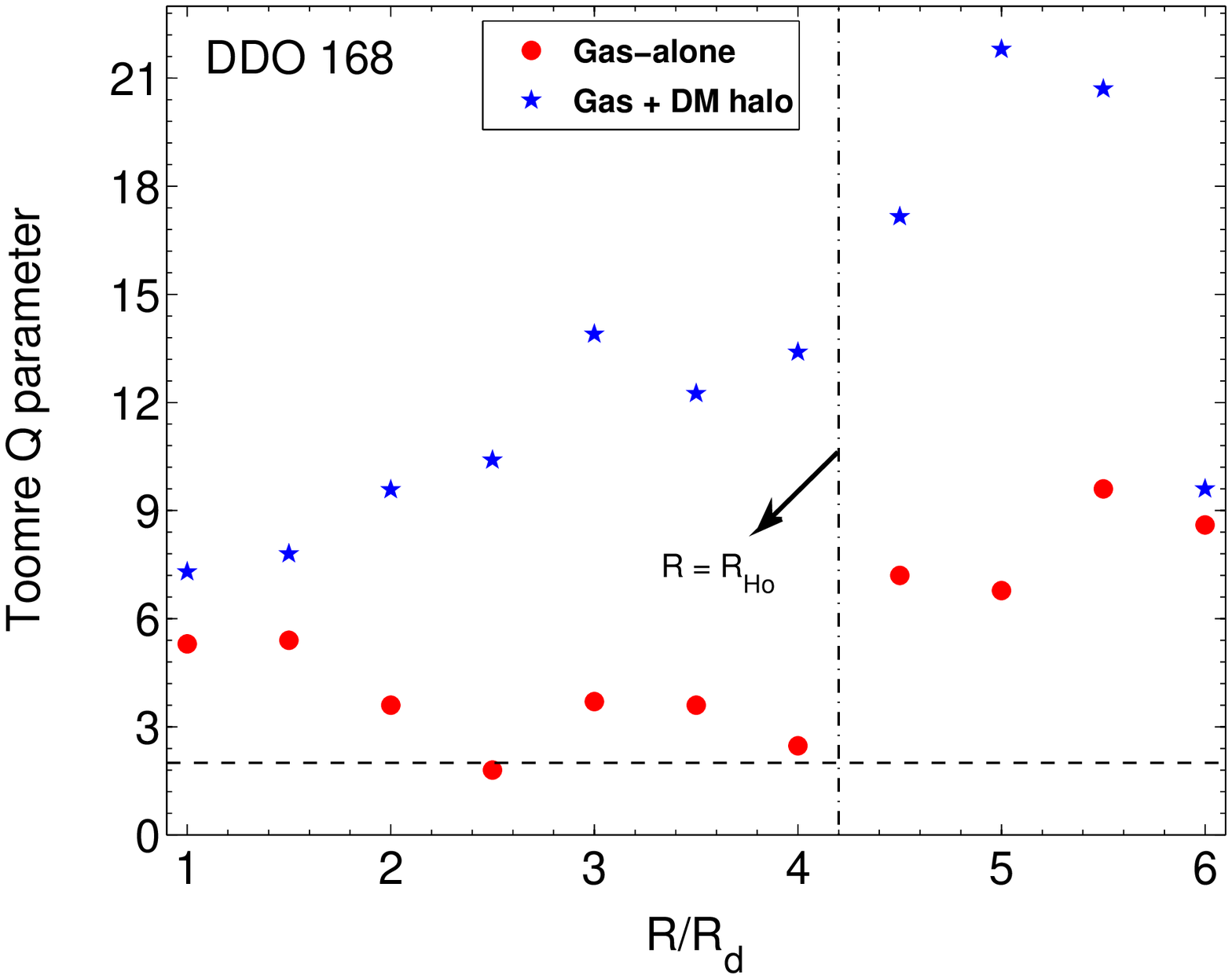}\par
\end{multicols}
\caption{Toomre $Q$ parameter for gas-alone and gas plus dark matter halo case, plotted as a function of radius for galaxies DDO~154, DDO~43, NGC~3741, and DDO~168.
The vertical line shown in each plot indicates the Holmberg radius ($R_{\rm Ho}$) which denotes the optical extent
for each galaxy. The horizontal dashed  line indicates $Q =2$ for each case which is taken to separate the region which allows swing amplification ($Q <2$) from region where it is suppressed ($Q>2$). The vertical dashed line denotes the Holmberg radius in each case.
Thus, this figure shows that the dark matter suppresses swing amplification at all radii, even inside of the optical radius in each case.}
\label{fig-q_plots}
\end{figure*}

To address this question, in this section we calculated the Toomre~$Q$ parameter for gas-alone case ($\kappa$ for only the gas component) and gas plus dark matter halo ($\kappa$ from the observed rotation curve) for a wide range of radii, ranging from the very  inner regions to the outermost possible region. This is shown in Fig.~(\ref{fig-q_plots}).

 We point out that, here we relaxed the assumption of flat rotation (used in Section 3.2 to calculate $\kappa$), and calculated the $\kappa$ values from the observed rotation curve using the standard definition $\kappa^2 = -4B(A-B)$, where A, B are Oort constants. 
When calculating the radial variation of $Q$, it is necessary to take account of $\kappa$ as dictated by the observed rotation curve. 
This difference in calculating $\kappa$ will give the Toomre $Q$ values slightly   different in these two cases. However, we checked that this difference is small, and will not alter the findings presented in the earlier subsections. 

The dashed horizontal line in each case denoted $Q=2$, which is taken to indicate the threshold so that $Q < 2$ permits swing amplification while it is damped for $Q > 2$, as is typically done \citep[e.g.][]{Too81}. However we caution that $Q = 2$ is an indicator but not a rigorous cut-off : 
 \citet{Too81} showed that for disks with {\it flat rotation curve}, $1\le X \le 3$ and $Q \le 2$ are the necessary and sufficient condition for swing amplification factors of more than a few \citep[also see][]{BT87}. However, we point out that the rotation curves of these dwarf galaxies are not strictly flat, therefore the above condition will not be exactly valid. 

 The vertical dashed line denotes the Holmberg radius, R$_{\rm Ho}$ in each case. We show next that the dark matter damps swing amplification even at radii inside of the optical radius in most cases.

 We note that gas surface density dominates over stellar surface density except in very central regions (of $\sim$ 1 $R_{\rm d}$), hence the application of one-component swing amplification treatment to gas in \S~2.2 is valid.

Here, we summarize the main results that we found from the study of the variation of Toomre $Q$ parameter as a function of radius, for each galaxy.\\

{\it{DDO~154 :}} The Toomre $Q$ parameter for the gas-alone case turned out to be less than $Q =2$, for all radii shown in Fig.~(\ref{fig-q_plots}), thus indicating that the gas disk is able to support the small-scale spiral features even within the optical disk. But the addition of dark matter produces a larger value of Toomre $Q$ parameter ($Q \sim 4$ being the typical values), and hence the system will no longer be able to produce the strong small-scale spiral features generated via swing amplification mechanism at all radii, even inside of the optical radius, R$_{\rm Ho}$.

{\it{DDO~43 :}} Here also for the gas-alone case, the values of Toomre $Q$ parameter are less than $Q = 2$ up to $R = 6.5 R_{\rm d}$ which also includes the optical disk and hence the system can display small-scale spiral structure. The addition of dark matter in the calculation increases the values of Toomre $Q$ parameter (see Fig.~(\ref{fig-q_plots})), thus the suppressing effect of dark matter halo on 
local spiral structure is seen at all radii.

{\it{NGC~3741 : }}The Toomre $Q$ parameter for the gas-alone case rises gradually to $Q \sim 4$ in the outermost point considered here. The addition of dark matter makes the values of Toomre $Q$ parameter as high as $11$ (a factor of $\sim$ 3 increment in Toomre $Q$ value) in the outer part (see Fig.~(\ref{fig-q_plots})). As indicated by the Toomre $Q$ for gas-alone case in the regions outside the optical disk, the gas disk still may display some weak small-scale spiral structure, the addition of dark matter rules out the possibility of having {\it small-scale} spiral features (but also see the \S~4.1). Again, the suppressing effect of dark matter halo on 
local spiral structure is seen at all radii, even inside of the optical radius, R$_{\rm Ho}$.

{\it{DDO~168 : }} The Toomre $Q$ parameter is found to be greater than $Q =2$ for all radii. The addition of dark matter makes the Toomre $Q$ parameter as high as $Q \sim 21$ in some radii. Hence, even though some weak small-scale spiral structure is possible to exist in the gas disk, the dominant dark matter completely rules out the possibility of having small-scale structure. Also, we note that the low disk surface density is another simultaneous possible reason (along with dominant dark matter halo) for not supporting any strong small-scale spiral features in the disk, as indicated by higher values of Toomre $Q$ parameter for gas-alone case.

Thus, we have shown that dark matter halo plays a vital role in preventing the strong small-scale spiral features in the outer disk as well as regions inside the optical disk. Also, for some cases, we found that the low disk surface density plays an important role along with the dark matter halo to prevent the small-scale spiral structure (e.g see cases of DDO~168 and NGC~3741 in Fig.~(\ref{fig-q_plots})).

\subsection{A counter-example : IC~2574}

\begin{figure}
    \centering
    \begin{minipage}{.5\textwidth}
        \centering
        \includegraphics[height=2.5in,width=3.5in]{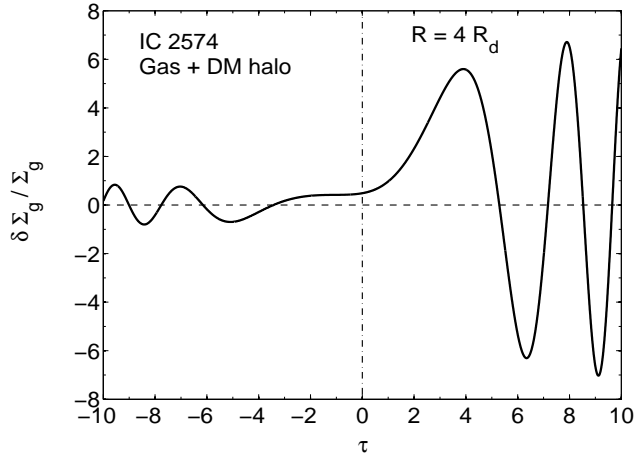}
    \end{minipage}
	\caption{IC~2574: variation in $\theta_{\rm g}$ = $\delta \Sigma_{\rm g}/\Sigma_{\rm g}$, the ratio of the perturbed gas surface density to the unperturbed gas surface density plotted as a function of $\tau$, dimensionless measure of time in the sheared frame at $R$ = 4 R$_{\rm d}$ ($Q$ = 1.2). The Toomre $Q$ value is calculated by taking the contributions of both gas and dark matter halo. The disk still allows a finite amplification, thus resulting in small-scale spiral features. The net fractional amplitude $\alpha \theta_{\rm g}$ is $\ll 1$ at all $\tau$, where $\alpha$ is a scale factor.}
    \label{fig-ic2574}
\end{figure}

 From Table~1, it is clear that the dark matter halo parameters for IC~2574 are quite different from the rest of the sample galaxies, i.e., it has a dark matter halo which is not dense and not compact. We point out that in the inner part ($R$ $\le$ 5 kpc) of this galaxy, baryons (stars and gas) are able to produce the observed rotation curve and only in the outer parts dark matter halo contribution takes over \citep[for details see fig.21 in][]{Oh08}. This is in a sharp contrast with the other galaxies in our sample where the dark matter halo is found to dominate from the innermost region. Therefore it is interesting to investigate the role of the dark matter halo on swing amplification process in such a different situation. We choose three radii, namely, 3 R$_{\rm d}$, 3.5 R$_{\rm d}$, and 4 R$_{\rm d}$. At these radii, the surface density values for $HI$ ($\Sigma_{HI}$) are 8, 6.5, and 4.9 M$_{\odot}$ pc$^{-2}$, respectively \citep{Wal08}. The Toomre $Q$ values for the gas plus dark matter halo case at radii 3 R$_{\rm d}$ and 3.5 R$_{\rm d}$ turn out to be less than one, indicating that the gas disk is unstable against the local axisymmetric perturbation and also is likely to host strong small-scale spiral structure. We note that the maximum rotation velocity is similar to those other galaxies, but due to a larger value of R$_{\rm d}$, positions of 3R$_{\rm d}$ and 3.5 R$_{\rm d}$ are further out as compared to the other galaxies, implying a lower value of $\kappa$ (as for a flat rotation curve $\kappa$ inversely proportional to the radius $R$); combinations of all these yield such a low value for Toomre $Q$. At $R$ = 4 R$_{\rm d}$, $Q$ is found to be 1.2, and it allows the system to have a finite swing amplification. This is shown is Fig.~\ref{fig-ic2574}.

Thus for IC~2574,  where the dark matter halo does not dominate  from the innermost radii, we find that 
the galaxy is likely to host small-scale spiral features. This could explain the spiral structure seen in IC~2574 which becomes visible when long exposure time is used \citep{dede64}.

\section{Discussion}

\subsection{ Swing amplification giving rise to local, material arms}

We note that in the literature the swing amplification process is applied as a generating mechanism for spiral 
structure in two scenarios.

In the first case, it is considered to generate local, material spiral features \citep[e.g.,][]{GLB65,JT66,Too81}. 
This local treatment is applicable at any point in the disk. In this paper, we have specifically treated local or small-scale swing amplified spiral arms (material arm) - as was first described in \citet{GLB65}, or in \citet{Too81}, and \citet{SC84}. This kind of spiral structure is shown to be chaotic in nature \citep{Sell11}. The local spiral features arising due to swing amplification are temporary by their very formation mechanism and show amplified density for a short time $\sim$ dynamical timescale. As a leading wave moves past the radial location, it can get swing amplified and then later as the pressure term dominates, then the trailing wave shows an oscillatory behaviour at a later time in the sheared frame \citep[e.g. see the description in][]{GLB65,Too81, Jog92}. Hence these have to be generated repeatedly. The origin of such recurring local features could be due to both the dissipation in gas and gas accretion, which will cool the disk and generate these features - as was shown by \citet{SC84}. Or these could arise due to perturbation caused by a giant gas complex \citep{JT66} or by a giant molecular cloud \citep{Don13}.

In the second case, swing amplification has often also been evoked to generate and maintain a
long-lived global spiral mode (e.g. \citet{Too81}, Chapter 3, 4; also see \citet{Mark74,Mark76}).
Here a more elaborate set-up is necessary and the wave transport across the corotation plus a 
feedback loop giving rise to a fresh supply of leading waves play a crucial role.
In the present paper we do not treat global spiral modes.

It may seem at first glance that since the swing amplification occurs over a finite time period, such material arms would vanish over a few rotation time-scales, thus leaving the gas disk featureless, without the dark matter halo playing a role in it.

However, we point out that in any linear perturbation analysis \citep[e.g. see][]{Rol77,BT87}, typically only the stability or the growth of the perturbation is considered without any explicit mention about what gives rise to such perturbation in the first place. This particular aspect of linear perturbation analysis was also followed in the studies of swing amplification in a galactic disk by \citet{GLB65,JT66,Too81}.

In this paper, we have tried to show that the gas disk in the dwarf irregular galaxies with extended HI disk is susceptible to swing amplification of a linear perturbation (as indicated by a lower Toomre $Q$ value) irrespective of its origin. However, the same gas disk in the presence of a dominant dark matter halo is no longer able to support the swing amplification (as reflected by a higher Toomre $Q$ value) due to the enhanced rotational stability provided by the dominant dark matter halo. Therefore, given the presence of all possible sources of perturbations, it is the set of disk and halo parameters and the resulting Toomre $Q$ value which decide whether the perturbation can grow in such a disk. 

 We note that gas-infall from outside is shown to be common and has significant dynamical effect on various processes in galaxy dynamics and secular evolution \cite[e.g. see][]{vanSa04,Com08,Fra08,SAN08,Com14}. Though we have not taken account of gas accretion as it is beyond the scope of our paper (and also the scope of classical analytic theories of swing amplification), the gas infall could be one such way to produce a fresh set of perturbations and subsequently to generate such spiral arms repeatedly.

\subsection{Occasional presence of spiral arms}

Our present analysis clearly shows that the dominant dark matter halo prevents almost completely the small-scale spiral structure by making the swing amplification process inefficient. However we note that one of our sample galaxies, namely, NGC~3741 shows a 
spiral arm in the disk, 
albeit it is faint and so in view of the results in our paper, its origin
remains a puzzle.

We point out that the swing amplification process stops being effective when $Q$ is greater than  2.5 and value of $X$ being greater than 3 \citep{Too81,SC84} and outside this range of parameter values the amplification will not be very strong \citep{BT87}. However it does not necessarily nullify the possibility of having occasional spiral features, as say triggered
by tidal encounters  \citep{ToTo72,BT87}.  The disk still can support occasional small-scale features although the higher values of $Q$ indicate that the response of the disk will not be very strong, thus making the spiral features weak \citep[for a detailed discussion, see \S~4.1 in][]{GJ14}. 

Some dwarf galaxies, which are not included in our sample, also show spiral structure. For example, the rotation curve decomposition of UGC~2259 shows that the behavior is typical of HSB galaxies \citep[see fig.9 in][]{CSV88}. It shows a grand-design spiral structure, although  faint: it is interesting to see that galaxies with such low luminosity (M$_B$ $\sim$ -16) are still susceptible to spiral instabilities \citep[for detailed discussion see \S~5 in][]{Ash92}.

 Also we caution the reader that here we are concerned about only late-type dwarf irregular galaxies 
which are rotationally supported.
Some of the early-type dwarf galaxies do show some spiral features \citep[e.g., see][]{Jer00,Gra03,Lis09}, but those are largely pressure-supported, and not rotationally-supported. Hence, they are dynamically different from the ones we consider in this paper.

Secondly note that two of our sample galaxies host a bar, e.g, DDO~154 and NGC~3741. 
However, the measured pattern speed of the bar is low, as in NGC~3741 \citep{Ban13}. This could be because of the dynamical friction due the dark matter halo which slows down the rotation speed of the bar significantly, as  proposed by \citet{DeSe98}. Bars can excite the transient spiral arms in the disk and it can heat up the disk as well, as was shown for HSB galaxies \citep{STT10}. The present paper shows that 
even when a bar exists, it seems unable to trigger strong spiral arms in our sample galaxies, perhaps these are suppressed by the dominant, dense and compact dark matter halo.

\subsection{Other issues}

Here we mention a few other points regarding this work.

First of all, galaxies with highly extended $HI$ distributions which are mainly studied here are not too common \citep{BCK05}, however so far no systematic study seems to have been done in the literature to see how common such galaxies are. Such a study ( Faint Irregular Galaxies GMRT survey 2 --FIGGS2) has now been started by \citet{Pat16}.
Therefore, an important question is: whether our finding is valid only for those dwarf irregular galaxies with extremely large $HI$ disks, or can it also be applicable to more typical dwarf irregular galaxies with moderately extended $HI$ disks?

In order to address this issue, it should first be noted that the size of $HI$ disk present in a galaxy sensitively depends on the limiting $HI$ column density, as the latter sets the limit up to which one can trace the $HI$ extent. 
For example, in NGC~3741, the measurement were done to the limiting $HI$ column density of 10$^{19}$ cm$^{2}$, while for DDO~154, NGC~2366 and IC~2574 taken from the THINGS sample \citep{Wal08}, the limiting value is typically 4 $\times$ 10$^{19}$ cm$^{-2}$. For LITTLE THINGS (from which DDO~168, DDO~43 are taken), in some of their samples (e.g., DDO~126, DDO~155) $HI$ column density is measured as low as a few times 10$^{17}$ cm$^{-2}$\citep[see fig.~5 in][]{Hun12}. For a detailed discussion of the size of the $HI$ disk see section 12.2 in \citet{GioHay88}. This choice affects the value of the outermost radius to which $HI$ is detected (see Table~1).

Further, in typical HSB galaxies, the optical radius is about 4--5 times the disk scalelength \citep{BM98}. This follows from the fact that the central luminosity is nearly constant and the outer
limit is set by the limiting magnitude of 26.5 mag arcsec$ ^{-2}$ \citep[see e.g.][]{vase81}; also for a detailed discussion see \citet{NJ03}. However in the dwarf irregular galaxies, the ratio can vary from 1 to 5 \citep{Swa99}. The reason is that the central value could be much lower while the outer limiting magnitude set by the sky brightness remains the same.
\citet{Swa99} showed that the ratio of the radial $HI$ extent to the optical size in dwarf irregular galaxies is $\sim 1.8$ similar to the ratio of $\sim 1.5$ for larger galaxies. We checked from the references used in Table~1 that, the ratio $R_{ HI}/ R_{\rm Ho}$ for DDO~154 and NGC~3741 is 4 and 8.3, respectively. For the other galaxies, namely, NGC~2366, DDO~43, and DDO~168, this ratio is 2.3, 2.2, and 2.8, respectively.

However, even the dwarfs that do not have very extended $HI$ distributions,  have a high $M/L$ ratio of $\sim 10$ \citep[e.g. see][]{CCF00}. Some of these dwarf galaxies do have a dense, compact halo (e.g., UGCA 442) where the dark matter halo dominates the rotation curve even at the inner radii.
 Thus the main result from this paper, namely, the dominant dark matter halo suppressing the generation of strong, small-scale spiral structure via swing amplification, is also valid for a much larger number of dwarf galaxies than the prototypical galaxies like DDO 154 and NGC 3741 that we have highlighted in our work.
However,
 the radial distance over which the dark matter halo dominates is not so large unlike the sample we have studied, so the cushioning influence
of the dark matter halo in preventing spiral structure, and resultant galaxy evolution, would not be as strong as in the case of sample galaxies studied in our work.

 We note that there are many other dwarf galaxies which have a lower density, non-compact halo (with $R_c/R_d > 3$), such as
the counter-example IC 2574 considered here, or HO II \citep[see e.g.][]{BJ13}. The dark matter halo in such galaxies is likely to play little role in the spiral structure formation in these.

Secondly, in any late-type galaxy stars and gas co-exist, and hence a more realistic approach would be to model the galactic disk as a gravitationally coupled two-fluid system and then do the non-axisymmetric perturbation analysis \citep[as done in][]{Jog92}. For the radii that we have considered here, $HI$ surface density greatly dominates over stars, e.g., for DDO~154, at $R$ = 6 R$_{\rm d}$, the stellar surface density is 0.015 M$_{\odot}$ pc$^{-2}$ as compared to a $HI$ surface density of 2.1 M$_{\odot}$ pc$^{-2}$. Therefore the system effectively reduces to a one-component system. This justifies our use of one-component formalism in \S~2.2.

\section{Conclusion}
In summary, we have carried out a 
 local, non-axisymmetric perturbation analysis for a sample of five late-type
dwarf irregular galaxies with extended $HI$ disks, and which have a dense and compact dark matter halo. 
  We show that when the gas disk is treated as an isolated system (with no dark matter included), the disk allows the growth of non-axisymmetric perturbations via swing amplification, but  the addition of a dominant dark matter halo in the analysis results in a higher rotational velocity and hence a higher Toomre $Q$ parameter which prevents the amplification almost completely. We further calculated the values of Toomre $Q$ parameter for a wide range of radii for these galaxies and showed that even in the regions inside the optical radius, the dominant dark matter halo yields  higher Toomre $Q$ values than the gas-alone case.
This naturally explains why these galaxies do not generally show strong small-scale spiral features despite being gas-rich.

 \citet{GJ14} showed that for the LSBs, where the stars dominate the baryonic disk, the dense and compact halo that dominates the disk at all radii suppresses the strong small-scale spiral features almost completely. 
{\emph {Thus, whether the larger baryonic contribution is in the form of stars or in the low velocity dispersion component $HI$ gas, in the region where the galaxy is dominated by dark matter halo, 
the disk will not be able to host small-scale spiral structure. }}

This suppression of spiral features inhibits further dynamical evolution, that would otherwise have been expected via angular momentum transport due to spiral features (see \S~1). This could be of interest in understanding early galaxy evolution, since in the hierarchical scenario of galaxy formation the smaller galaxies form first and then merge to form larger galaxies.
The  dwarf irregular galaxies, being  small, low-metalicity, gas-rich systems; are believed to be the present-day analogs of the high-redshift unevolved galaxies. Hence understanding the dynamical evolution of dwarf irregular galaxies could possibly give some insight in to the early evolution of high-redshift galaxies.

\bigskip
\noindent {\bf Acknowledgements}: We thank the anonymous referee for a thoughtful report which has helped to improve the paper. We thank Elias Brinks for making available tabular data \citep[from][]{Wal08} on
gas distribution in DDO 154, NGC 2366 and IC 2574.
CJ would like to thank the DST, Government of India for support via a
J.C. Bose fellowship (SB/S2/JCB-31/2014).

\end{document}